\documentclass[12pt]{iopart}
\pdfoutput=1 

\usepackage{graphicx,epstopdf}  
\expandafter\let\csname equation*\endcsname\relax
\expandafter\let\csname endequation*\endcsname\relax
\usepackage{amsmath}
\usepackage{graphicx}
\usepackage{amsfonts}
\usepackage{color}
\usepackage{dcolumn}   
\usepackage{bm}        
\usepackage{amssymb}   
\usepackage{hyperref}
\usepackage{amsmath}
\usepackage{amsfonts}
\usepackage{amssymb}

\begin{document}

	\newcommand{\angstrom}{\text{\normalfont\AA}}
	\newcommand{\braket}[3]{\bra{#1}\;#2\;\ket{#3}}
	\newcommand{\projop}[2]{ \ket{#1}\bra{#2}}
	\newcommand{\ket}[1]{ |\;#1\;\rangle}
	\newcommand{\bra}[1]{ \langle\;#1\;|}
	\newcommand{\iprod}[2]{ \langle#1|#2\rangle}
	\newcommand{\intl}[2]{\int\limits_{#1}^{#2}}
	\newcommand{\logt}[1]{\log_2\left(#1\right)}
	
	\newcommand{\mc}[1]{\mathcal{#1}}
	\newcommand{\mb}[1]{\mathbb{#1}}
	\newcommand{\cx}[1]{\tilde{#1}}
	\newcommand{\nn}{\nonumber}
	\newcommand{\la}{\langle}
	\newcommand{\ra}{\rangle}
	\newcommand{\blang}{\big \langle}
	\newcommand{\brang}{\big \rangle}
	
	\newcommand{\p}{\partial}
	\newcommand{\cmnt}[2]{\textbf{\#\#}{\color{#1}#2}\textbf{\#\#}}
	
	\newcommand{\flap}{\mb{L}_{\bar{\kappa}}}
	\newcommand{\flapfinite}{\mb{L}}
	\newcommand{\fcurr}{\mb{A}}
	\newcommand{\flapFull}{|\Delta|^{3/4}}
	\newcommand{\tdir}{f}
	\newcommand{\mzeta}{\chi}
		\newcommand{\momsym}{\delta}
	\def\be{\begin{equation}}
	\def\ee{\end{equation}}
	\def\bea{\begin{eqnarray}}
	\def\eea{\end{eqnarray}}
	
	\newcommand{\eqa}[1]{\begin{align}#1\end{align}}
	\newcommand{\mbf}[1]{\mathbf{#1}}
	\newcommand{\iu}{{i\mkern1mu}}
	\newcommand{\widesim}[2][1.5]{
		\mathrel{\overset{#2}{\scalebox{#1}[1]{$\sim$}}}
	}
	
	\newcommand{\sina}{\alpha}
	\newcommand{\HTheta}{\theta}
	\newcommand{\genpower}{\beta}
	\newcommand{\cosa}{\phi}
	\newcommand{\RN}[1]{
		\textup{\uppercase\expandafter{\romannumeral#1}}%
	}
	
	\title[Dynamical correlations in the hard particle gas]{Dynamical correlations of conserved quantities in the  one-dimensional equal mass hard particle gas}

	\author[1]{Aritra Kundu$^{1,2}$,  Abhishek Dhar$^1$, Sanjib Sabhapandit$^2$}
	\address{$^{1}$International Centre for Theoretical Sciences, Tata Institute of Fundamental Research,  Bengaluru 560089, India\\$^2$Raman Research Institute, Bangalore 560080, India}
	\ead{aritrak@icts.res.in, sanjib.sabhapandit@gmail.com, abhishek.dhar@icts.res.in,  }
	\vspace{10pt}
	\begin{indented}
		\item[]\today
	\end{indented}

	\begin{abstract}
We study a gas of point particles with hard-core repulsion in one dimension where  the particles move freely in-between elastic collisions. We prepare the system with a uniform density on the infinite line. The velocities  $\{v_i; i \in \mathbb{Z} \}$ of the  particles are chosen independently from a thermal distribution. Using a mapping to the non-interacting gas, we analytically compute the  equilibrium spatio-temporal  correlations $\la v_i^m(t) v_j^n(0)\ra$ for arbitrary integers $m,n$. The analytical results are verified with microscopic simulations of the Hamiltonian dynamics. The correlation functions have  ballistic scaling, as expected in an integrable model.
	\end{abstract}



\section{Introduction}

The study of equilibrium spatio-temporal correlations in one-dimensional systems of interacting particles with Hamiltonian dynamics is useful in understanding   their transport properties. Non-integrable  systems have a few conserved quantities and  spatio-temporal correlation functions in one dimension show universal behavior that can be understood within the framework of fluctuating hydrodynamics of the conserved quantities \cite{VanBeijeren2012,Spohn2014,Narayan2002}. In contrast, for integrable systems, which are characterized by a macroscopic number of conservation laws, the spatio-temporal correlations are non-universal and very limited results are known.  Some progress has recently been made in understanding transport in integrable systems through the approach of generalized hydrodynamics for these macroscopic number of conserved quantities, see for example \cite{Castro2016,Doyon2018,Doyon2018Soliton,Bastianello2018,Pavlov2003} and references therein.

One of the simplest examples of an interacting particle  system is the hard particle gas (HPG), where particles undergo elastic binary collisions, while in between the collisions, they move freely with constant velocity. The elastic collisions conserve energy and momentum. In general, when the masses of the particles are not all equal, the system is expected to be non-integrable.  A particular case, where the masses of successive particles are taken to alternate between two values,  has been studied both in the equilibrium and non-equilibrium setups, and is known to exhibit anomalous transport with the thermal conductivity diverging with system size \cite{Casati1986,Garrido2001,Grassberger2002,Dhar2001,Hurtado2016,Chen2014}.  In contrast, in the equal mass HPG, the particles  simply exchange their velocities during  the elastic collisions, which makes the system integrable.  Evidently, the conserved quantities in this system are $\sum_i v_i^n$ for integer $n$. It was shown by Jepsen \cite{Jepsen1965} that one can effectively map the equal mass HPG to a gas of non-interacting particles and many exact results, such as tagged particle equilibrium velocity correlations, could be obtained. Some extensions to the case of hard rods were obtained in  \cite{Lebowitz1967}. In recent work \cite{Sabhapandit2015,Hegde2014,Roy2013}, considerable simplification of this approach was used to obtain exact results on properties of tagged particle displacements and velocity correlations.  More recently \cite{Kundu2016},  the correlations of energy, momentum and  stretch was computed numerically in Toda chain and the hard particle case was studied as a limiting case of Toda potential.

In this paper, following the methods in \cite{Sabhapandit2015}, we  compute dynamical correlations for all conserved quantities which, as noted above, are simply given by various powers of the velocity. 
Specifically,  we considered a set of $2N+1$ particles of unit masses that are initially distributed uniformly inside a one-dimensional box of length $2L$. The initial  velocitity of each particle is chosen independently from the Maxwell-Boltzmann distribution at temperature $T$.  The ordered particles have  positions  and velocities given by $\{ x_i, v_i\}$ for $i=1,2\ldots,2N+1$.   
We are interested in computing general spatio-temporal correlation functions,  defined as 
\eqa{\langle v_i^m(t); v_j^n(0) \rangle   =\left[ \langle v_i^m(t)v_j^n(0) \rangle - \langle v_i^m(t)\rangle \langle v_j^n(0) \rangle \right]/\bar{v}^{m+n} ,~~{\rm with}~ \bar{v}=\sqrt{T},  \label{eq:corrfunct}}
where $m,n$ are positive integers and  $i,j$ are taken to be particles in the bulk, and the average is over initial configurations  chosen from the equilibrium Gibb's distribution at temperature $T$ and uniform particle density.
In the thermodynamic limit of $N \to \infty, L \to \infty$ while keeping the density $\rho=N/L$ constant, the correlation function depends only on the relative position of the two particles, $r = i-j$.

Our main results include the following explicit asymptotic ($t \to \infty$) form for the velocity correlations, in terms of the scaling variable $l=r/(\rho \bar{v} t)$:
\eqa{  \rho \bar{v} t \langle v^m_{r}(t); v^n_0(0) \rangle = (l^m - \momsym_m)(l^n - \momsym_n) f(l),}
where $f(\omega)=e^{-\omega^2/2}/\sqrt{2 \pi}$ and $\momsym_n=\int_{-\infty}^\infty d\omega \omega^n f(\omega)$.
From these, in particular, we extract the  correlations  for stretch, momenta and energy as
\eqa{
\rho \bar{v} t\la s_{r}(t) s_0(0) \ra   &= \frac{1}{\rho^2}  f(l) \\
\rho \bar{v} t\la v_r(t) v_0(0) \ra  &= {\bar{v}^2} l^2 f(l) \\
\rho \bar{v} t \la e_r(t);e_0(0 \ra   &= \frac{\bar{v}^4}{4} \left( l^4-2 l^2 +1
\right) f(l)~. 
}

The paper is organized as follows: In  Sec.~\eqref{sec:1system},  we define the model precisely and explain the idea of computing the correlations exactly using the mapping to the non-interacting system. Then in Sec.~\eqref{sec:asymp} we present results for the correlations which are obtained in the asymptotic limit of large time. We also verify the results from simulations of the microscopic system.  Finally we conclude in Sec.~\eqref{sec:3conc}.

\section{Main steps of the calculation}
\label{sec:1system}

\begin{figure}
	\centering
	\includegraphics[width=0.7\linewidth]{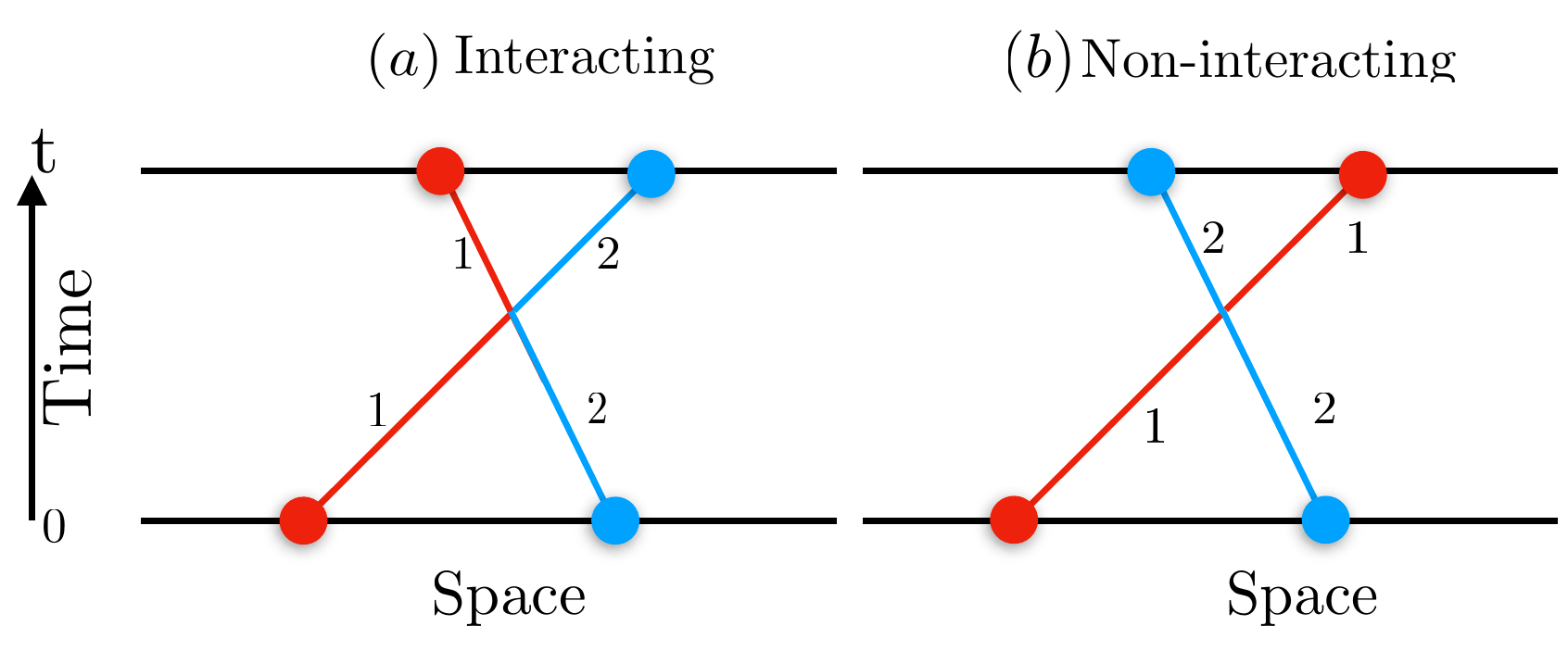}
	\caption{An interacting hard-point particles system (a) can be constructed from a non-interacting system by exchanging tags (colors) when two trajectories cross (b).}
	\label{fig:interacting-non-interacting}
\end{figure}

We consider a gas of $2N+1$ ordered particles with positions $\{x_i\}$, with $i=1,\ldots,2N+1$,  initially distributed uniformly in the  interval $[-L,L]$.  The  velocities $\{v_i \}$ are  chosen from the Gibbs distribution  $P(\{v_i\}) = \prod_i [e^{-v_i^2/2T}/\sqrt{2 \pi T}]$. Since we are interested only in the thermodynamic limit and in bulk properties, we do not need to include a confining wall.  
The particles move freely with constant velocity between elastic collisions with the neighbors. Under a collision the particles simply exchange their velocities.  This allows us to make a mapping from interacting system  of particles  [Fig. \ref{fig:interacting-non-interacting}(a)], where the particles collide with each other, to a  system of non-interacting particles  [Fig. \ref{fig:interacting-non-interacting} (b)], where the particles evolve independently and pass through each other, and we exchange labels whenever particles cross.  In fact,  this non-interacting picture allows one to evolve the system directly to any final time and the corresponding trajectory in the case where the particles are undergoing collisions can be obtained by relabeling the tags of the particles.
The probability of obtaining the  trajectories in the non-interacting picture is same as that of the interacting system. 

This mapping to the non-interacting gas was used by Jepsen \cite{Jepsen1965} to obtain an exact 
solution for velocity-velocity autocorrelation functions in the hard-particle gas. A simpler approach was recently proposed in \cite{Sabhapandit2015,Hegde2014,Roy2013} to obtain two particle distribution, tagged particle statistics and also a particular case of velocity correlations with $m=n=1$ in Eq.~\eqref{eq:corrfunct}.

In the non-interacting picture,  a particle at $x$ with velocity $v$ travels to  $y$ at time $t$ such that  $y =  x+ vt = x + \sigma_t v/ \bar{v} $ where we have defined $\sigma_t = \bar{v} t$.   For a particle with velocity chosen from 
the Maxwell distribution, the  single particle propagator giving the probability of  finding the particle at $y$ at time $t$ is then given by 
\eqa{G(y,t|x,0) =& \int_{-\infty}^\infty  dv \delta(y-x- \sigma_t v/ \bar{v})  \frac{e^{-v^2/2\bar{v}^2
	}}{\sqrt{2 \pi \bar{v}^2}}   = \frac{1}{\sigma_t} f\left(\frac{y-x}{\sigma_t}\right), \label{eq:propagator} }
where the function  $f(\omega) = e^{- {\omega^2}/{2}}/{\sqrt{2 \pi}}$. Note that most parts of the derivations given below go through for arbitrary scaling function $f(\omega        )$ with finite moments. 

Our  strategy  is to compute the correlations of velocity  via the two-time joint  distribution function $P(x,j,0;y,k,t)$ defined as the probability of the $j^{th}$ (ordered) particle being at $x$ at time $t=0$ and the $k^{th}$ (ordered) particle being at $y$ at time $t$. 
Following \cite{Sabhapandit2015},  this joint PDF can be readily computed using the single particle propagator and through the mapping to non-interacting particles. We now give the   details.

\subsection{Computation of  the joint probability distribution of two particles}
\begin{figure}
	\centering
	\includegraphics[width=0.7\linewidth]{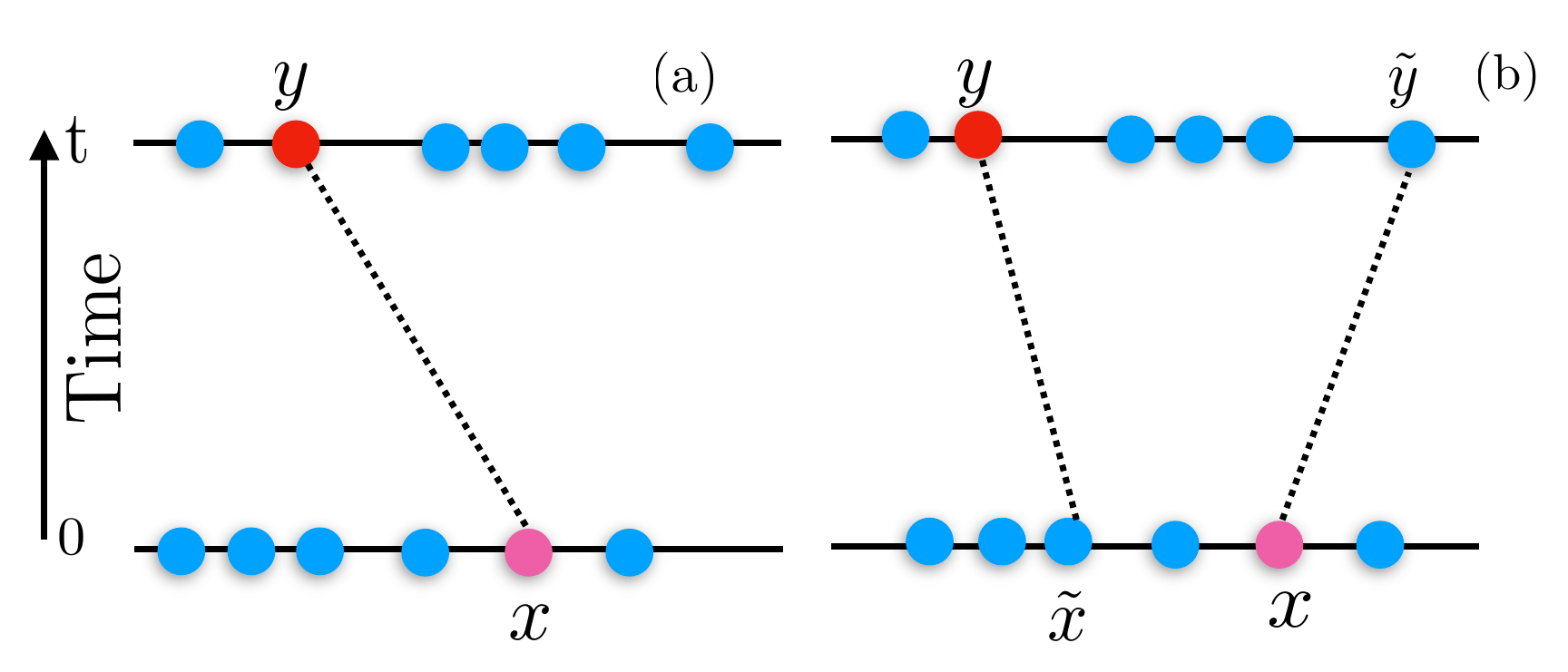}
	\caption{Illustration of the two different possibilities in the non-interacting picture, that contibute to the propagator $P(x,j,0,y,k,t)$ for the interacting system. Here we consider $6$ particles  with $j=5$, $k=2$. The  two possibilities are then (a) the $j$th particle, at time $t=0$, becomes the $k$th particle at time $t$ and (b) a different particle ($i=3$) becomes the $k$th particle at time $t$.}
	\label{fig:propagator}
\end{figure}

In terms of the non-interacting gas picture, the joint probability 
density, $P(x,j,0;y,k,t)$, of the $j^{th}$ (ordered) particle being at $x$ at time $t=0$ and the $k^{th}$ (ordered) particle being at $y$ at time $t$, has two  independent contributions $P_1(x,j,0;y,k,t)$ and $P_2(x,j,\tilde{x},0;y,k,\tilde{y},t)$ that are defined below (see Fig.~\ref{fig:propagator}). 
\begin{itemize}
	\item[(i)] In the non-interacting gas picture, $P_1(x,j,0;y,k,t)$ is the probability that $j^{th}$ particle,  which is at $x$ at time $t = 0$, becomes the $k^{th}$ particle at $y$ at time $t$ [shown in Fig.~\ref{fig:propagator}(a)]. 
	This is obtained by  picking one of the non-interacting particles at random at time $t = 0$, with probability $ (2N+1)/(2L) $,  evolving with the propagator from $(x,0)$ to $(y,t)$ and  multiplying by the probability that  this is  the $j^{th}$ particle at time $t=0$ and   the $k^{th}$ particle  at time $t$. We thus get
	\be
	P_1(x,j,0;y,k,t) = \frac{2N+1}{2L} G(y,t|x,0) F_{1N} (x,j,0;y,k,t)~,
	\ee
	where $F_{1N} (x,j,0;y,k,t)$ is the probability that there are $j-1$ particles to the left of $x$ at $t=0$ and $k-1$ particles to the left of $y$ at time $t$. 
	The explicit form of $F_{1N}$ can be obtained by combinatorial arguments \cite{Sabhapandit2015} and is given in the Appendix.
	\item[(ii)] Similarly, for the non-interacting gas, $P_2(x,j,\tilde{x},0;y,k,\tilde{y},t)$ is the probability that the $j^{th}$ particle at position $x$ at $t=0$ becomes another particle at $\cx{y}$ at time $t$, and some other particle at $\tilde{x}$ at time $t=0$ becomes the $k^{th}$ particle at position $y$ at time $t$ [shown in Fig.~\ref{fig:propagator}(b)].
	In this case,
	two particles are chosen randomly at $t=0$, with probability $(2N+1)(2N)/(2L)^2$, and are evolved with the propagator for the transition $(x,\tilde{x},t=0) \to (\tilde{y},y,t)$. Finally we need to multiply this with the probability, $F_{2N} (x,j,\tilde{x},0;y,k,\tilde{y},t)$, that there are  $j-1$ particles to the left of $x$ at $t = 0$ and $k-1$ particles to the left of $y$ at time $t$, given that there is a particle at $\tilde{x}$ at $t=0$ and a particle at $\tilde{y}$ at time $t$. Combining these, we get
	\be
	P_2(x,j,\tilde{x},0;y,k,\tilde{y},t) = \frac{(2N+1)(2N)}{(2L)^2} G(y,t|\tilde{x},0)G(\tilde{y},t|x,0) F_{2N} (x,j,\tilde{x},0;y,k,\tilde{y},t)~.
	\ee
	The explicit form of $F_{2N}$ can be obtained by combinatorial arguments \cite{Sabhapandit2015} and is given in the Appendix.
\end{itemize}

\subsection{Relating the joint probability distribution to correlations}

The  contribution to velocity correlations in the interacting system comes from two separate processes corresponding to the two joint probability distributions from the previous section. Hence we obtain
\eqa{ \mathcal{F}^{(m,n)} (k,j,t) = \la v_k^m(t)v_j^n(0)\ra/\bar{v}^{(m+n)}  &=  \underbrace{\la v_k^m(t)v_j^n(0)\ra_1/\bar{v}^{m+n}}_{[\RN{1}]}  + \underbrace{\la v_k^m(t)v_j^n(0)\ra_2 /\bar{v}^{m+n}}_{[\RN{2}]} \nn ,}
where $[\RN{1}]$ and $[\RN{2}]$ can be computed as follows.

(i) The first contribution comes when the $j^{th}$ particle at $x$ at time $t = 0$ becomes the $k^{th}$ particle at $y$ at time $t$. In the non-interacting picture, these two particles are correlated and have the same velocity, i.e.,   $v_k(t)/\bar{v} = v_j(0)/\bar{v} =(y-x)/\sigma_t$. Multiplying this with the appropriate probability  $ P_1(x,j,0;y,k,t)$ and integrating over all possible initial and final positions, we get the first contribution to velocity correlations
\eqa{
	[\RN{1}] &= \intl{-\infty}{\infty} dx \intl{-\infty}{\infty} dy ~ {\left(\frac{y-x}{ \sigma_t}\right)}^{m+n} P_1(x,j,0;y,k,t) \nn,\\
	&= \rho \sigma_t  \intl{-\infty}{\infty} \frac{dx}{\sigma_t} \intl{-\infty}{\infty} \frac{dy}{\sigma_t} ~ {\left(\frac{y-x}{ \sigma_t}\right)}^{m+n} f\left(\frac{y-x}{ \sigma_t}\right)F_{1N}(x,j,y,k),  \label{eq:vcnapprox1}
}
where we have replaced $(2N+1)/(2L)$ with the density $\rho$ in the thermodynamic limit. 

(ii) The second contribution is when the $j^{th}$ particle at $x$ with velocity $ v_j(0)/\bar{v} =(\tilde{y}-x)/\sigma_t$ at $t=0$ becomes another particle at $\tilde{y}$ time $t$,   and some other particle at $\tilde{x}$ with velocity $ v_k(t)/\bar{v} =(y-\tilde{x})/\sigma_t$  at time $t=0$ becomes the $k^{th}$ particle at $y$ at time $t$. Multiplying this with the appropriate probability  $P_2(x,j,\tilde{x},0;y,k,\tilde{y},t)$ and integrating over 
all position variables, we get
\eqa{ 
	[\RN{2}] 	&=  \intl{-\infty}{\infty} dx \intl{-\infty}{\infty} dy \intl{-\infty}{\infty} d\tilde{x} \intl{-\infty}{\infty}  d\tilde{y}~ {
		\left(\frac{\tilde{y}-x}{ \sigma_t}\right)}^m {\left(\frac{y-\tilde{x}}{ \sigma_t}\right)}^n
	P_2(x,j,\tilde{x},0;y,k,\tilde{y},t), \nn \\
	&= (\rho \sigma_t)^2  \intl{-\infty}{\infty} \frac{dx}{\sigma_t} \intl{-\infty}{\infty} \frac{dy}{\sigma_t} \intl{-\infty}{\infty} \frac{d\cx{x}}{\sigma_t} \intl{-\infty}{\infty} \frac{d\cx{y}}{\sigma_t}  ~ 
	{\left(\frac{\tilde{y}-x}{ \sigma_t}\right)}^m {\left(\frac{y-\tilde{x}}{ \sigma_t}\right)}^n  \nn \\
	&\hskip 4cm \times	f\left(\frac{\cx{y}-x}{ \sigma_t}\right)f\left(\frac{y-\cx{x}}{ \sigma_t}\right)   F_{2N} (x,j,\tilde{x},0;y,k,\tilde{y},t), \label{eq:vcnapprox2}
}
where again we have used the thermodynamic limit density $\rho$.

\section{Asymptotic results  in the thermodynamic limit }
\label{sec:asymp} 
\subsection{Computation of velocity correlations}
Next we use the explicit forms of $F_{1N}, F_{2N}$  given in the Appendix.  One can  perform the integrals over $\cx{x},\cx{y}$  in the second term $[\RN{2}]$ in Eq.~(\ref{eq:vcnapprox2}). We also make a change of integration variables from  $x,y$  to the variables $z = (x-y)/\sigma_t$ and $\bar{z} = (x+y)/\sigma_t$, in Eq.~\eqref{eq:vcnapprox1} and Eq.\eqref{eq:vcnapprox2}. After some  algebra,  we finally obtain the following expression for the velocity correlations, for $j$ in the bulk: 
\eqa{	C^{(m,n)}(r,t)&=\label{eq:Cnm1}
	\mathcal{F}^{(m,n)} (j,j+r,t) = [\RN{1}]+[\RN{2}]\\ &=\rho\sigma_t 
	\intl{-\infty}{\infty} dz
	\intl{-\infty}{\infty}  \frac{d\bar{z}}{2}
	\intl{-\pi/2}{\pi/2}  \frac{d\phi}{\pi}
	\intl{-\pi}{\pi}  \frac{d\theta}{2\pi}
	D(z,\theta,\phi)   \nn \\ & \times e^{-2N(1-\cos \phi)} e^{\iu \rho \sigma_t(-\bar{z}\sin \phi + z \sin \theta)}e^{i(\phi r - \theta r)}e^{-2 \rho \sigma_t Q(z) (1-\cos \theta)} ,\nn \\
	{\rm where}~~ Q(z) &= z \intl{0}{z} dw~ f(\omega) + \intl{z}{\infty} d\omega~ \omega f(\omega),\\
		D(z,\theta,\phi) &= (\rho \sigma_t)^{-1} z^{m+n}f(z) + \Delta_1^m(z)\Delta_2^n(z)e^{-\iu \phi} \nn\\
	 & + \Delta_1^n(z)\Delta_2^m(z)e^{\iu \phi}  + \Delta_2^m(z)\Delta_2^n(z)e^{\iu \theta}+ \Delta_1^m(z)\Delta_1^n(z)e^{-\iu \theta},} 
and the $\Delta_{1,2}^p$ are related to moments of the propagator
\eqa{ &\Delta_1^p(z) = \intl{z}{\infty} d\omega ~\omega^p f(\omega)
	,~~\Delta_2^p(z) = \intl{-\infty}{z} d\omega ~\omega^p f(\omega)~.
}
Since in Eq.~\ref{eq:Cnm1} the $\phi$ term comes with a factor of $N$, a saddle point analysis reveals that the major contribution comes from $\phi=0$. We do a Taylor expansion around $\phi=0$, up to second order, first perform the resulting Gaussian integral in $\phi$ and then perform the resulting Gaussian integral in the variable  $\bar{z}$. This  leads to following simpler form
\eqa{
	{C}^{(m,n)} (r,t) = \rho\sigma_t 
	\intl{-\infty}{\infty} dz
	\intl{-\pi}{\pi}  \frac{d\theta}{2\pi}
	D(z,\theta,0) e^{- \rho \sigma_t [2Q(z)(1-\cos \theta) - \iu z \sin \theta]}e^{-\iu \theta r}.\label{Fv4}
}
We define  a new scaling variable $l=r/(\rho \sigma_t)$ and rewrite the above equation as
\eqa{
	{C}^{(m,n)} (r=\rho \sigma_t l,t)  = \rho\sigma_t 
	\intl{-\infty}{\infty} dz
	\intl{-\pi}{\pi}  \frac{d\theta}{2\pi}
	D(z,\theta,0) e^{- \rho \sigma_t I(z,\theta)},
	\label{eq:Cnm2}
}
where, $  I(z,\theta) =  [2Q(z)(1-\cos \theta) - \iu z \sin \theta+\iu \theta l]$.  Now we are interested in the large time (hence $\rho\sigma_t >> 1$) behaviour and so we again use saddle point methods. Our strategy is to find $(z^*,\theta^*)$, where the function $I(z,\theta)$ is minimum, and expand both the functions $I$ and $D$ in Eq.~(\ref{Fv4}) around this minimum. The minimum is obtained by solving the equations
\eqa{
	2  Q(z^*)\sin \theta^*-i  z^* \cos \theta^* + i l =0, \nonumber \\
	2  Q'(z^*) (1-\cos \theta^*)-i   \sin \theta^* =0.
}
whose solution is $z^*=l, \theta^*=0$.
We define new scaling variables $(w,u)$ through  $z = l + w \sqrt{\epsilon}$ and $\theta = u \sqrt{\epsilon}$, where $\epsilon =1 /\rho \sigma_t$, which gives
\eqa{\label{Cexp}
	\frac{ I(l+w\sqrt{\epsilon},u\sqrt{\epsilon})}{\epsilon} &=  (u^2 Q(l)-i u w) + C_1\epsilon^{1/2} +C_2 \epsilon + \mc{O}(\epsilon^{3/2})~,  \\
	{\text{with}}~~
	C_1 &=  \left(u^2 w Q'(l)+\frac{1}{6} i l u^3\right),~~~
	C_2 = -\frac{1}{12} u^2 \left(-6 w^2 Q''(l)+u^2 Q(l)-2 i u w\right), \nn}
and
\eqa{\label{Dexp}
&D( l + w\sqrt{\epsilon},u\sqrt{\epsilon})~=~\delta_m \delta_n + 
 D_1\epsilon^{1/2} +D_2 \epsilon + \mc{O}(\epsilon^{3/2})~, \\
{\rm with}~&D_1 = i u \left[ \Delta_2^m(l) \Delta_2^n(l) - \Delta_1^m(l) \Delta_1^n(l) \right], \\
&D_2 = 	  f(l) l^{m+n}\nn -\frac{1}{2} u^2  \left[ \Delta_2^m(l) \Delta_2^n(l) + \Delta_1^m(l) \Delta_1^n(l) \right] 
+ i u w f(l) \left(l^n \delta_m+l^m \delta_n\right),  \\
{\rm and}~& \momsym_p = \Delta_1^p(z) + \Delta_2^p(z)  =
	\int_{-\infty}^\infty \omega^p f(\omega)d\omega. \label{delta_m}
}
 For $f(\omega)=e^{-\omega^2/2}/\sqrt{2\pi}$, we have  $\momsym_m=  [1+(-1)^m] (m-1)!!/2$,  where $!!$ denotes the double factorial.
Consequently, we get the following series expansion in powers $\sqrt{\epsilon}$:
\eqa{\label{Aexp}
D(l + w\sqrt{\epsilon},u\sqrt{\epsilon}) e^{-C_1 \epsilon^{1/2} -C_2 \epsilon+\ldots} = \sum_{k=0}^\infty a_k(u,w) (\sqrt{\epsilon})^{k}~,
}
where $a_k(u,w)$ are polymonials in $u$ and $w$. 
Using Eqs.~(\ref{Cexp},\ref{Dexp},\ref{Aexp})  in Eq.~\eqref{eq:Cnm2}, we then  get 
\eqa{
	&{C}^{(m,n)} (r=\rho \sigma_t l,t)  = 
	\intl{-\infty}{\infty} du
	\intl{-\infty}{\infty}  \frac{d w}{2\pi} e^{-u^2 Q(l)- i u w} \sum_{k=0}^\infty a_k(u,w) (\sqrt{\epsilon})^{k}. 	\label{eq:Cnm3}
}
Finally, we perform  Gaussian integrations over the variables $u$ and $w$ to get
\eqa{{C}^{(m,n)} (r=\rho \sigma_t l,t) = \momsym_m \momsym_n+(l^m - \momsym_m)(l^n - \momsym_n) f(l)/\rho \sigma_t  +\mc{O}(\epsilon^{3/2}),
	\label{eq:corr}}
where $\momsym_n$ is defined in \eref{delta_m}. The scaled correlation function after subtracting off the mean is then finally given, to leading order, by
\eqa{\rho \sigma_t\langle v^m_{r=\rho\sigma_t l}(t); v^n_0(0) \rangle  = (l^m - \momsym_m)(l^n - \momsym_n) f(l). \label{vvcorr}}

\subsection{Computation of other correlations}
Here we compute some other correlations which are of interest from the point of hydrodynamics. We define the stretch variable $s_l=q_{l+1}-q_l$.
To  compute the stretch correlations,  we note that
\eqa{
	\la s_r(t) s_0(0) \ra  &= \la [ q_{r+1}(t) -q_r(t) )(q_r(0) -q_0(0)] \ra \nn \\
	~~~~&= -  \left[ \la q_{r+1}(t) q_0(0) \ra - 2 \la q_r(t) q_0(0) \ra + \la q_{r-1}(t) q_0(0) \ra \right] \nn\\
	&\approx -\partial_r^2 \la(q_r(t)q_0(0) \ra~,
}
where we have used the translation symmetry of the system. 
Now taking two time derivatives  gives
\begin{align}
	\partial^2 _t \la s_r(t) s_0(0) \ra =  -\partial^2_r \la(v_r(t)v_0(0) \ra , 
\end{align}
where we used the results $\la v_r(t) q_0(0)\ra=\la v_r(0) q_0(-t)\ra$ and $\la v_r(0) v_0(-t)\ra=\la v_r(t) v_0(0)\ra$,  following from time-translation invariance. The stretch correlation can be written in terms of velocity correlations as follows:
\bea
\la  s_r(t) s_0(0) \ra  = \la  s_r(0)\ra \la s_0(0) \ra-\int_{0}^{t} dt' \int_{0}^{t'}dt''  \partial^2_r \la v_r(t'')v_0(0) \ra,
\eea
where we used the fact that $\la v_r(0) s_0(0) \ra=0$. 
This finally gives 
\be
\la  s_r(t); s_0(0) \ra= \la  s_r(t) s_0(0) \ra-\la  s_r(0)\ra \la s_0(0) \ra = \frac{1}{\rho^3  \sigma_t}  \frac{e^{-\frac{1}{2}({\frac{r}{\rho \sigma_t}})^2}}{\sqrt{2 \pi}}~.
\ee
The energy correlations are easily expressed in terms of velocity correlations. 
\eqa{
	\la e_r(t);e_0(0) \ra &=  \la e_r(t)e_0(0) \ra - \la e_r(t) \ra \la e_0(0) \ra 
	=\frac{1}{4} \la [v^2_r(t) - \la v^2_r(0) \ra] [ v^2_0(0) -\la v^2_0(0)\ra ] \ra  \nn~.
}
Using the known form of the velocity correlations from \eqref{vvcorr}, we then get
\be
\la e_r(t);e_0(0) \ra = \frac{\bar{v}^4}{\rho \sigma_t} \left[ \left( \frac{r}{\rho \sigma_t}\right)^4 - 2\left( \frac{r}{\rho \sigma_t}\right)^2 + 1 \right] f\left(\frac{r}{\rho \sigma_t}\right)~.
\ee

Following  arguments similar to those in \cite{Sabhapandit2015},  we now show the asymptotic exact results obtained above can be understood from a heuristic argument. Since the initial
velocities are chosen independently for each particle, the contribution to the correlation
function $\la v^m_r(t);v^n_0(0)\ra$ is non-zero only when the velocity of the $r$th particle at time $t$ is the same as that of the zero-th particle at time $0$.
The initial velocity distribution of each particle is chosen from a Maxwell distribution ${e^{- v^2/2\bar{v}^2}}/{\sqrt{2 \pi}\bar{v}}$, with $\bar{v}^2=k_B T$. The velocity correlation function is thus approximately given by
\eqa{
\la v^m_r(t);v^n_0(0)\ra & \approx 
\int_{-\infty}^\infty dv \frac{[v^m-\la v^m\ra]}{\bar{v}^m} \frac{[v^n-\la v^n\ra]}{\bar{v}^n} \frac{1}{\rho t} \delta \left(v-\frac{r}{\rho t}\right) \frac{1}{\bar{v}}f\left(\frac{v}{\bar{v}}\right) \\
&=\frac{1}{\rho \sigma_t}  (l^m - \momsym_m)(l^n - \momsym_n) f(l),
}
which reproduces \eqref{vvcorr}.

In the next section we  verify this result from simulations with the microscopic dynamics of the hard particle gas.

\subsection{Numerical verification with Hamiltonian evolution of the hard-particle gas}
\label{sec:4Numerical}
We now present results from numerical simulations of the hard particle gas. 
In our simulations we considered a gas of hard point particles with   density $\rho=1$, moving on a ring. We choose the initial condition from an equilibrium distribution at temperature $T$, i.e.,  we distribute the particles uniformly in space with unit density and  velocity of each particle is independently chosen from the distribution $e^{-v^2/(2T)}/\sqrt{2 \pi T}$ . 
The mapping to the independent particle picture means that the time-evolution of this system can be done very efficiently. Basically, starting from any given initial condition,  we evolve the non-interacting gas up to time $t$. In order to get the actual positions, we can get the correct tag of the interacting particle by simply sorting their final positions and taking into account the effect of periodic boundaries. We then compute the correlations $ \la v^m_l(t);v^n_0(0)\ra$ by taking averages over initial conditions. In our simulations we took $1000$ particles and took averages over  $10^8$ initial conditions. 
\begin{figure}
	
	\centering

	\includegraphics[width=\linewidth]{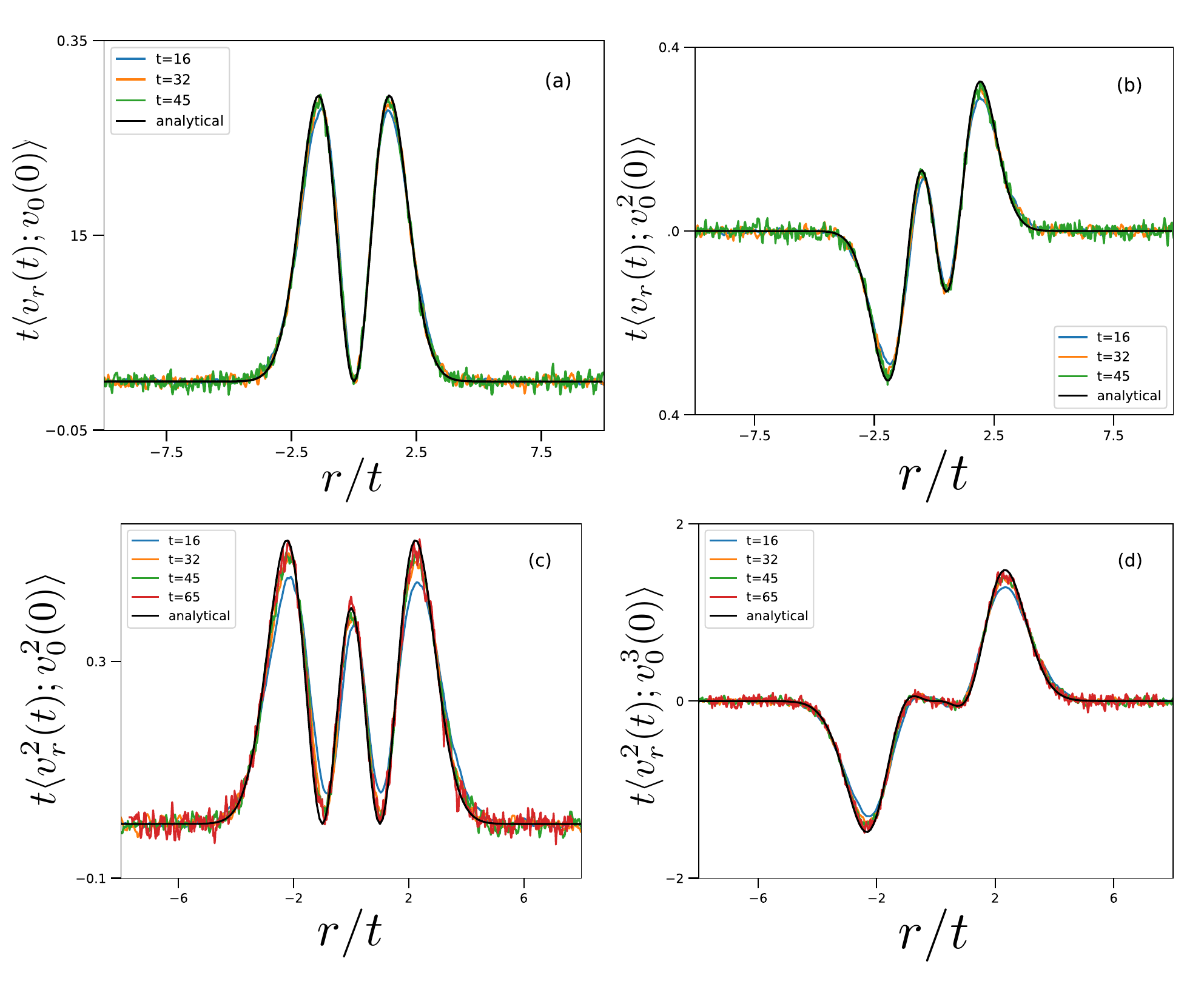}
	\caption{Scaled velocity correlation functions at different times, for various choices  of $m,n$ as obtained from simulations of the HPG.  The black curves are from the analytical result as given by Eq.~\eqref{eq:corr}. The curves are for $t << L$. The parameters in the simulation are $N=1001$, $L=1000$ and temperature $T=1$.}
	\label{fig:corr}
\end{figure}
In Fig.~(\ref{fig:corr}) we plot the scaled correlation functions  $t \la v^m_{r}(t);v^n_0(0)\ra $  (corresponding to ballistic scaling), obtained from the microscopic simulations and  compared  with the corresponding theoretical 
predictions from Eq.~\eqref{vvcorr},  for several choices of $m,n$.  We find very good agreement between simulations and theory, even at relatively early times. 
At very short times there is a small deviation from the theory and this is expected, since the theory makes predictions for long time behaviour, well after the transient dynamics.
On the other hand, finite size effects would show up at times $t \gtrsim \mc{O}(L)$. 
The peak of the correlation functions at time $t$,  for odd values of $m,n$, is given by $r_{peak} \approx  \bar{v}t\,  \sqrt{m+n} $. In contrast to this, the sound velocity  for the HPG is given by $\sqrt{3}\bar{v}$. Using this, we estimate that finite size effects  would show up in correlation functions of order $m+n$, at times approximately $t^* \sim L/(\bar{v}\sqrt{m+n})$.

\section{Conclusion}\label{sec:3conc}
Using the mapping of the HPG to a non-interacting gas, we have 
computed exact  long-time spatio-temporal correlation functions of arbitrary powers  of velocities. This gives us  correlations of all conserved quantities in the HPG. We have shown how other correlations such as the ``stretch'' variable can be obtained.   We have  verified our analytic results through direct simulations of the HPG using a very efficient numerical scheme, which again relies on  the mapping to non-interacting particles. 
While we obtained the correlations to leading order in $1/(\rho \bar{v} t)$, our approach can be readily extended to systematically compute the corrections, which can sometimes be important.  Indeed, our approach here has been used in earlier work to compute tagged particle correlations, where the higher order corrections are relevant\cite{Roy2013,Hegde2014,Sabhapandit2015}.    
The hard particle gas is in the same class as the harmonic chain, as an example of a ``non-interacting'' integrable model \cite{Spohn2018} but the  dynamics is  non-Gaussian and so the computation of correlations is somewhat more non-trivial than for the harmonic chain. We expect that our results will provide a bench-mark to test predictions from generalized hydrodynamics of integrable systems.

\section{Acknowledgments}
 AD and SS would like to acknowledge the support from the Indo-French Centre for the promotion of advanced research (IFCPAR) under Project No. 5604-2.

\section{Appendix}
\subsection{Computation of $F_{1N}$ and $F_{2N}$}\label{app:f1f2}
The expressions for $F_{1N}(x,y,j,k,t)$ and  $F_{2N}(x,y,\cx{x},\cx{y},j,k,t)$  were explicitly computed in \cite{Sabhapandit2015}, using combinatorial arguments. Here  we give the explicit expressions:
\eqa{
	F_{1N}(x,y,j,k,t) =& \intl{-\pi/2}{\pi/2} \frac{d\phi}{\pi}	\intl{-\pi}{\pi} \frac{d\theta}{2\pi} [H(x,y,\theta,\phi,t)]^{2N} e^{-\iu \phi (2N+2 -k + j )} e^{-\iu \theta (k - j) },\\
	F_{2N}(x,y,\cx{x},\cx{y},j,k,t) =& \intl{-\pi/2}{\pi/2} \frac{d\phi}{\pi}	\intl{-\pi}{\pi} \frac{d\theta}{2\pi} [H(x,y,\theta,\phi,t)]^{2N-1		} e^{-\iu \phi (2N+2 -k + j )} e^{-\iu \theta (k - j)} e^{-\iu \phi \chi_1} e^{-\iu \theta \chi_2} ,
}
where $\chi_1,\chi_2$ are defined as follows
\begin{enumerate}
	\item $\cx{x} < x$ and $\cx{y} < y$ , then $\chi_1 =1 $ and $\chi_2 = 0$,
	\item $\cx{x} < x$ and $\cx{y} > y$ , then $\chi_1 =0 $ and $\chi_2 = 1$,
	\item $\cx{x} > x$ and $\cx{y} < y$ , then $\chi_1 =0 $ and $\chi_2 = -1	$,
	\item $\cx{x} > x$ and $\cx{y} > y$ , then $\chi_1 =-1 $ and $\chi_2 = 0$.
\end{enumerate}
The function $H(x,y,\theta,\phi,t) $ is defined as
\eqa{ H(x,y,\theta,\phi,t) &= p_{++}(x,y,t)e^{\iu \phi} + p_{--}(x,y,t)e^{-\iu \phi} + p_{+-}(x,y,t)e^{\iu \theta} + p_{-+}(x,y,t)e^{-\iu \theta} , \nn \\
	&= 1 - (1- \cos \phi) (p_{++} + p_{--}) + \iu \sin \phi (p_{++} - p_{--}), \nn \\
	&- (1- \cos \theta) (p_{+-} + p_{-+}) + \iu \sin \theta (p_{+-} - p_{-+}), \label{eq:Hpolar}}
where  $p_{-+}(x,y,t)$ is defined as the probability that a single non-interacting particle is to the left of $x$ at time $t=0$ and to the right of $y$ at time $t$, and $p_{+-},p_{--}$ and $p_{++}$ are defined similarly. Their explicit forms are
\begin{eqnarray}
	p_{-+}(x,y,t) = \frac{1}{2L}\intl{-L}{x}dx'\intl{y}{L} dy'~G(y',t|x',0), \\
	p_{+-}(x,y,t) = \frac{1}{2L}\intl{y}{L}dx' \intl{-L}{x}dy'~G(y',t|x',0), \\
	p_{++}(x,y,t) = \frac{1}{2L}\intl{x}{L}dx'\intl{y}{L} dy'~G(y',t|x',0), \\
	p_{+-}(x,y,t) = \frac{1}{2L}\intl{-L}{x}dx' \intl{-L}{x}dy'~G(y',t|x',0)~.
\end{eqnarray}
We note that $p_{-+}+p_{+-}+p_{--}+p_{++}=1$. 
We can explicitly find the expressions for $p_{\pm\pm}$ using the exact propagator Eq.~\eqref{eq:propagator}. In the large $N,L$ limit, expanding in $1/L$, we get
\begin{eqnarray}
	p_{-+} &=& \frac{\sigma_t}{2L}\left(-\frac{-z}{2} + Q(z)\right)	+ \mathcal{O}(1/L^2), \nn \\
	p_{+-} &=& \frac{\sigma_t}{2L}\left(\frac{-z}{2} + Q(z)\right) + \mathcal{O}(1/L^2) ,\nn \\
	p_{++} &=& \frac{1}{2} + \frac{\sigma_t}{2L}\left( -\frac{\bar{z}}{2} -Q(z)\right) + \mathcal{O}(1/L^2) ,\\
	p_{--} &=& \frac{1}{2} + \frac{\sigma_t}{2L}\left( \frac{\bar{z}}{2}-Q(z)\right) + \mathcal{O}(1/L^2) \nn ,
\end{eqnarray}
where $z = (x-y)/\sigma_t$ and $\bar{z} = (x+y)/\sigma_t$ and \eqa{Q(z) = z \intl{0}{z} dw~ f(w) + \intl{z}{\infty} dw~ wf(w).} Now, substituting these asymptotic expressions of $p_{\pm\pm}$ in Eq.~\eqref{eq:Hpolar} for large $N$, keeping only the dominant terms, one finds
\eqa{ [H(x,y,\theta,\phi,t)]^{2N} = e^{-2N(1-\cos \phi)}e^{\iu \rho \sigma_t \bar{z} \sin \phi}e^{-2\rho\sigma_tQ(z)(1-\cos \theta)}e^{\iu \rho \sigma_t z \sin \theta}.}

		\section{References}

\providecommand{\newblock}{}

		\appendix
		
	\end{document}